\documentstyle[12pt,aasms4,flushrt]{article}

\received{}
\accepted{}

\begin{document}

\title{Jet stability and the generation of superluminal and stationary
components}

\author{Iv\'an Agudo\altaffilmark{1}, Jos\'e-Luis G\'omez\altaffilmark{1},
Jos\'e-Mar\'{\i}a Mart\'{\i}\altaffilmark{2}, Jos\'e-Mar\'{\i}a
Ib\'a\~nez\altaffilmark{2}, Alan P. Marscher\altaffilmark{3}, Antonio
Alberdi\altaffilmark{1}, Miguel-Angel Aloy\altaffilmark{4},
Philip E. Hardee\altaffilmark{5}}

\altaffiltext{1}{Instituto de Astrof\'{\i}sica de Andaluc\'{\i}a, CSIC,
Apartado 3004, 18080 Granada, Spain. ivan@iaa.es; jlgomez@iaa.es;
antxon@iaa.es}

\altaffiltext{2}{Departamento de Astronom\'{\i}a y Astrof\'{\i}sica,
Universidad de Valencia, 46100 Burjassot (Valencia), Spain.
jose-maria.marti@uv.es; jose.m.ibanez@uv.es}

\altaffiltext{3}{Institute for Astrophysical Research, Boston University, 725
Commonwealth Avenue, Boston, MA 02215, USA. marscher@bu.edu}

\altaffiltext{4}{Max-Planck-Institut f\"ur Astrophysik, Karl-Schwarzschild-Str.
1, D-85748 Garching, Germany. maa@mpa-garching.mpg.de}

\altaffiltext{5}{Department of Physics \& Astronomy, The University of Alabama,
Tuscaloosa, AL 35487, USA. hardee@athena.astr.ua.edu}

\begin{abstract}

  We present a numerical simulation of the response of an expanding
relativistic jet to the ejection of a superluminal component. The simulation
has been performed with a relativistic time-dependent hydrodynamical code from
which simulated radio maps are computed by integrating the transfer equations
for synchrotron radiation. The interaction of the superluminal component with
the underlying jet results in the formation of multiple conical shocks behind
the main perturbation.  These trailing components can be easily distinguished
because they appear to be released from the primary superluminal component,
instead of being ejected from the core. Their oblique nature should also
result in distinct polarization properties. Those appearing closer to the core
show small apparent motions and a very slow secular decrease in brightness,
and could be identified as stationary components. Those appearing farther
downstream are weaker and can reach superluminal apparent motions. The
existence of these trailing components indicates that not all observed
components necessarily represent major perturbations at the jet inlet; rather,
multiple emission components can be generated by a single disturbance in the
jet. While the superluminal component associated with the primary perturbation
exhibits a rather stable pattern speed, trailing components have velocities
that increase with distance from the core but move at less than the jet
speed. The trailing components exhibit motion and structure consistent with
the triggering of pinch modes by the superluminal component. The increase in
velocity of the trailing components is an indirect consequence of the
acceleration of the expanding fluid, which is assumed to be relativistically
hot; if observed, such accelerations would therefore favor an
electron-positron (as opposed to proton rest-mass) dominated jet.


\end{abstract}

\section{Introduction}

  The computation of the synchrotron radio emission (G\'omez et al. 1995,
1997; Hughes, Duncan \& Mioduszewski 1996; Komissarov \& Falle 1996) from
time-dependent relativistic hydrodynamical codes (Mart\'{\i}, M\"uller, \&
Ib\'a\~nez 1994; Duncan \& Hughes 1994; Koide, Nishikawa, \& Mutel 1996) has
proven to be a powerful method to understand the physics of jets in AGNs and
microquasars improving upon previous idealized analytical calculations
(Marscher \& Gear 1985, Hughes et al. 1985). Superluminal components have been
studied using such formulations by introducing perturbations in the jet inlet
and analyzing the resulting shock waves and their subsequent evolution along
the jet (G\'omez et al. 1997; Mioduszewski et al. 1997). Fluctuations in the
internal structure of the shocked material may be amplified by time delay
effects, resulting in a knotty brightness distribution of the superluminal
component associated with the shocked plasma (G\'omez et al. 1997).

  In this {\it Letter} we explore the instabilities produced in the jet
resulting from the evolution of strong perturbations generated at the jet
inlet. In particular, we describe a mechanism, which should be common in jets,
that generates multiple superluminal as well as quasi-stationary components.

\section{Generation of Traveling Perturbations in Relativistic Jets}

  Our study relies on a relativistic, axially-symmetric jet model obtained by
means of a high-resolution shock capturing scheme to solve the equations of
relativistic hydrodynamics in cylindrical coordinates. The code is the same as
that used by G\'omez et al. (1997). Details of the code and its performance
(equations, finite--difference approximation, and testing) can be found in
Mart\'{\i} et al. (1997, and references therein).

The steady jet model is the same as model PM in G\'omez et al. (1997),
computed with a spatial resolution in both radial and axial directions of 8
cells/$R_b$ (where $R_b$ is the beam radius at the injection position), but
twice as long (covering 400x10 $R_b$). It corresponds to a pressure-matched,
diffuse ($\rho_b/\rho_a=10^{-3}$), relativistic ($\Gamma_b=4$) beam with Mach
number $M_b=1.69$. Here $\rho$ is the proper rest-mass density and $\Gamma$ is
the bulk Lorentz factor; subscripts $a$ and $b$ refer, respectively, to
atmosphere and beam; values correspond to the injection position. By allowing
the jet to propagate through an isothermal atmosphere with a decreasing
pressure gradient, we have induced a small opening angle in the jet ($\sim$
0.3$^\circ$).

\subsection{Dynamics of the main perturbation}

We concentrate our attention on the evolution of the flow after the
introduction of a square-wave perturbation at the jet inlet. This perturbation
consists of an increase in the bulk Lorentz factor from the quiescent value of
$\Gamma_b=4$ to $\Gamma_p=10$, and an increase in pressure by a factor of
two. The perturbation lasts for a time $\tau_p=0.75 R_b/c$ (where $c$ is the
speed of light), after which the jet inlet was set to the quiescent values
both in Lorentz factor and pressure. The fluid piles up in front of the
velocity perturbation, creating a fast shocked state. Given the large initial
shock velocity of $\sim$0.995 ($\Gamma=10$), in order to determine relative
variations in this velocity of the order of 10\% (for example, to discern
between a shock Lorentz factor of 10 and, say, 11), the relative error in the
calculation of the distance traveled by the shock must be smaller than
1/1000. This precision can only be obtained by studying the evolution of the
shock over more than 1000 computational cells. Given the fact that the shock
front is spread over 3-4 cells and the computational domain covers 3200 cells,
only mean shock speeds (of 0.995 $c$) can be obtained. In general, the Lorentz
factor of the leading shock should be larger than that of the pre-shock flow
($\sim 4$ at $ z = 0$; $\sim 12$ at $z = 400 R_b$). This shocked state is
followed by a more slowly moving (mean speed of 0.973 $c$) rarefaction where
the fast flow separates from the slower upstream flow. This difference in
speeds stretches the perturbation along the axis. Any similar strong {\it
supersonic} perturbation should lead to a similar shock/rarefaction structure.

The passage of the main perturbation triggers a local pinch instability
that propagates behind the main perturbation, leading to the formation of a
series of conical ``{\it trailing shocks}'' following the main perturbation
(Fig. 1). Shocked and rarefied parts of the main perturbation and variation in
the beam radius can be seen in the figure. The trailing structures are spaced
by $\sim 8~R_b$ for $15 < z/R_b < 55$ and $\sim 20~R_b$ for $105 < z/R_b <
180$.

Although the type of perturbation introduced at the jet inlet is somewhat
arbitrary, the qualitative results we obtain are not a function of the
particular perturbation chosen; formation of trailing shocks is a general
consequence of the propagation of strong perturbations through jets (see also
G\'omez et al. 1997, where trailing shocks appear in the hydrodynamical
simulations for a different jet inlet perturbation). Further research would be
of interest to quantify the physics of the trailing shocks depending on the
type of perturbation introduced, as well as the jet/external medium
hydrodynamical properties.

A local solution of the dispersion relation for the pinch modes of this
relativistic jet and a computation of the pinch mode structure (see Hardee et
al.\ 1998; Hardee 2000) indicate that observed features are primarily related
to triggering of the first pinch body mode. Features in Figure 1 in the range
$15 < z/R_b < 180$ have spacings between the longest unstable wavelength
$\lambda^{\it l}$ and the shorter maximally unstable wavelength $\lambda^*$ of
the first pinch body mode. The wavelengths
$\lambda^{\it l}$ and $\lambda^*$ increase from 13
$R_b$ to 125 $R_b$ and from 4 $R_b$ to 19 $R_b$, respectively, as $z$
increases from 0 to 250 $R_b$.

\subsection{Evolution of the Trailing Shocks}

  The observed time evolution of the trailing shocks can be followed in Figure
2, which displays a space-time diagram of the Lorentz factor distribution in
the jet. Trailing shocks emerge from the rarefactions that follow the main
perturbation, and are generated over a range of velocities and separations.
The velocities are significantly smaller than the velocity of the main
perturbation (and the bulk flow speed) and increase with distance from the
inlet, as do the separations. Table 1 summarizes the positions, separations,
and speeds of the trailing components for different observer's times. In
particular, consider time $t_{obs} = 60$ when the shock spacing from A to H ranges
from $\sim 17~R_b$ to $\sim 76~R_b$ and the shock velocity ranges from $\sim
0.13c$ to $\sim 0.69c$ for locations from $z/R_b = 12.7$ to $z/R_b =
290.9$. The separation and speed are explained by rapid passage of the main
perturbation, which triggers the first pinch body mode at a wavelength whose
group velocity is comparable to the rarefaction speed, $v_{gp} \sim 0.973~c$.
This wavelength ranges from $\lambda \sim 23~R_b$ at $z/R_b \sim 50$ to
$\lambda \sim 60~R_b$ at $z/R_b \sim 250$.  Subsequently, the perturbation
slows to the wave speed, $v_{ph}$, of this wavelength. The wave speed ranges
from $\sim 0.2~c$ at $z/R_b \sim 50$ to $\sim 0.6~c$ at $z/R_b \sim 250$.  The
wave speed is low for perturbations created with wavelength near $\lambda^{\it
l}$ (occurs at small $z$) and higher for perturbations created with wavelength
above $\lambda^{\it l}$ (occurs at larger $z$).  Another result is that
perturbations created at small $z$ (relatively short wavelength) accelerate as
they move to larger $z$ (see Fig.\ 2) where their wave speed is higher.

\section{Emission results}

  We have computed the synchrotron radiation from the hydrodynamical model
discussed in the preceding section following the procedure described in
G\'omez et al. (1997) and references therein. Fig. 3 shows the computed total
intensity image obtained for an optically thin observing frequency. Several
components are seen, corresponding to the trailing conical shocks of Figs. 1
and 2, and labeled accordingly. Typical high-frequency
VLBI observations provide dynamic
ranges of the order of 200:1 (defined as the intensity ratio of the peak to
the lowest believable feature) or larger, depending on the frequency of
observation and array used. All components seen in Fig. 3 are above this
limit, and therefore are expected to be detected by actual VLBI observations.

  Figure 4 displays the mean intensity along the observed jet in a space-time
diagram. Trailing components have been identified across epochs, and their
mean intensity peak is traced as a function of observer's time. Almost all the
trailing shocks in Fig. 2 can be identified with radio components in Fig. 4
and have been labeled accordingly. In Fig. 4 we can distinguish a strong
component associated with the main shock, followed by a jet region with low
emission corresponding to the rarefaction that follows the main
shock. Trailing components are observed to emerge upstream from this dip in
emission. Therefore, trailing components can be easily distinguished from those
generated at the jet inlet: they emanate not from the core, but rather seem to
appear spontaneously downstream, created in the wake of strong
perturbations. The main component is observed to move with a superluminal
apparent speed that slowly decreases from $\sim 7$ to $\sim 6 c$. Note that
from these velocities we can infer a Lorentz factor for the hydrodynamical
component between 7.3 and 6.1 (for a viewing angle of $\theta=10^\circ$),
smaller than the value of $\sim 10$ estimated from the hydrodynamics. This
apparent discrepancy may be explained by considering the time delays that
severely affect the leading component, stretching its size in the observer's
frame by a factor of $\beta\sin\theta/(1-\beta\cos\theta)$. Small changes in
the brightness distribution from the front to the back of the component would
then lead to a lag of the centroid behind the shock front and therefore to a
slower value of its proper motion as derived from VLBI images than that
expected from the velocity of the leading shock.

  As observed in Fig. 4, the flux densities and apparent motions of the
trailing components depend strongly on the distance from the core at which
they are generated, reflecting the hydrodynamics of the secondary
features. Those components appearing close to the core show subluminal motions
as well as flux densities that decrease very slowly with time. Components
appearing farther downstream show progressively larger apparent motions
(episodically superluminal) as well as weaker flux densities, rendering their
detection more difficult. Table 2 compiles positions and speeds, in the
observer's frame, of the trailing components at selected observer's times. The
values of $z_{obs}$ for each component in Table 2 can be readily compared with
those of $z$ in Table 1 once projection effects have been removed ($z_{obs}=z
\sin 10^\circ$). All the distances of emission components match positions of
peaks in Lorentz factor within an error smaller than 16\%, except for
components I and J. A similar comparison can be made between the mean apparent
speeds in the intervals $t_{obs}$=15--30, 30--60 and the true speeds in Table
2, although in this case the agreement is within a 30\%. The higher velocities
of components I and J result in longer time delays, which stretches their
emitting volume in the observer's frame, thus rendering the determination of
their positions more difficult.

\section{Discussion and Conclusions}

  Our relativistic hydrodynamic simulations show that the passage of a strong
perturbation along a jet triggers instabilities caused by pressure mismatches
between the jet and external medium, leading to the formation of conical
trailing recollimation shocks. The existence of these features has been
ignored in previous analyses in which analytical square-wave shock
approximations have been used. Computation of the synchrotron emission reveals
that these trailing components should be detectable with present VLBI arrays.

  Our simulations show that trailing components should appear to emerge in the
wake of strong components instead of ejected from the core. They are generated
with a wide range of apparent speeds, from almost stationary (near the core)
to superluminal -- although slower than the main leading feature -- farther
down the jet. They are conical (more generally oblique in three-dimensional
models) shocks, and hence should exhibit polarization properties that depend
on the viewing angle (Cawthorne \& Cobb 1990) and are different from those of
the main components. Evidence for the existence of trailing components have
been found in the radio galaxy 3C~120, whose jet contains a standing component
whose flux density is observed to be enhanced after the passage of a very
strong superluminal component (G\'omez et al. 2000). Recent high resolution 7
mm VLBA observations (Jorstad et al. 2001) have revealed the existence of
steady components close to the core in a large number of $\gamma$-ray bright
blazars; these may be associated with trailing components.

  The superluminal component corresponding to the main perturbation in our
simulation exhibits a rather constant pattern speed, even though the
underlying fluid velocity is observed to accelerate significantly. Therefore,
the observation of the proper motions of superluminal components may not
detect any actual acceleration of the jet fluid. However, our simulations show
that trailing components should appear to accelerate, reflecting acceleration
of the fluid. If such accelerations are not observed, this would indicate that
the fluid is not relativistically hot, as expected for an electron-positron
plasma. In this case, the internal energy of the jet could be inferred to have
an energy density dominated by the rest mass of nonrelativistic protons.

  From our analysis we conclude that jets and, by implication, the accretion
process onto the central black hole can be steadier than previously thought,
while still producing highly variable emission and prolific production of
superluminally moving knots. Multiple components in radio maps can be produced
by a single ejection event.

  The mechanism of production of these trailing components is a general
characteristic of nonsteady supersonic jet flows, i.e.,
not specific to relativistic dynamics. It could be applied to other
astrophysical scenarios such as outflows from young stellar objects where very
knotty jets are found with series of components moving at very different
speeds from those inferred for the fluid flow (Xu, Hardee \& Stone 2000).

  The spacing and acceleration of the trailing components can be
understood as resulting from the triggering of pinch modes by the main
perturbation.  Pinch wavelengths with group velocity comparable to the speed
of the rarefaction associated with the main perturbation are excited but then
decelerate to the phase speed of the dominant wavelength. The excited
wavelength is longer and the wave speed is higher at larger distance as an
indirect consequence of the acceleration of the expanding jet.

\begin{acknowledgements}

  This research was supported by Spanish Direcci\'on General de
Investigaci\'on Cient\'{\i}fica y T\'ecnica (grants PB97-1164,-1432), by NASA
Astrophysical Theory grant NAG5-3839, and by the Fulbright commission for
collaboration between the USA and Spain. M. A. Aloy acknowledges financial
support from the guest program of the Max-Planck-Institut f\"ur
Astrophysik and the fellowship number EX0022566499 from the Spanish 
Ministerio de Educaci\'on y Ciencia. P. Hardee acknowledges support from 
the National Science Foundation through grant AST-9802955 to the University 
of Alabama. P. Hardee and J. L. G\'omez acknowledge the Aspen Center for 
Physics where some of this work was performed.

\end{acknowledgements}

\begin{deluxetable}{crrrrrrrrrrr}
\tablecolumns{12}
\tablewidth{0pc}
\tablecaption{Positions, separations, and speeds of the trailing shocks 
              in the source frame at selected observer's times.\label{tab2}}
\tablehead{
  \multicolumn{1}{c}{ } &
  \multicolumn{4}{c}{$z (R_b)$} &  
  \multicolumn{1}{c}{ } &  
  \multicolumn{3}{c}{$\Delta z (R_b)$} &
  \multicolumn{1}{c}{$v (c)$ } &
  \multicolumn{2}{c}{$\bar{v} (c)$} \nl \cline{2-5} \cline{7-9}  \cline{11-12} 
  \multicolumn{1}{c}{Comp.} &
  \multicolumn{1}{c}{$15^a$} &
  \multicolumn{1}{c}{$30^a$} &
  \multicolumn{1}{c}{$50^a$} &
  \multicolumn{1}{c}{$60^a$} &
  \multicolumn{1}{c}{ } &
  \multicolumn{1}{c}{$15^a$} &
  \multicolumn{1}{c}{$30^a$} &
  \multicolumn{1}{c}{$60^a$} &
  \multicolumn{1}{c}{$15^a$} &
  \multicolumn{1}{c}{$15-30^a$} &
  \multicolumn{1}{c}{$30-60^a$} }
  
\startdata

A&   6.6 &   8.0 &  11.1 &  12.7& &     &      &       & 0.25 & 0.08 & 0.13 \nl
B&  23.2 &  25.4 &  28.4 &  30.1& &16.6 & 17.4 &  17.4 & 0.30 & 0.13 & 0.13 \nl
C&  38.5 &  42.1 &  50.8 &  49.7& &15.3 & 16.7 &  19.6 & 0.25 & 0.20 & 0.20 \nl
D&  57.7 &  63.8 &  78.3 &  80.0& &19.2 & 21.7 &  31.3 & 0.50 & 0.29 & 0.29 \nl
E&  75.6 &  89.4 & 106.9 & 116.7& &17.9 & 25.6 &  36.7 & 0.85 & 0.48 & 0.48 \nl
F& 103.4 & 116.7 & 144.3 & 158.7& &27.8 & 27.3 &  42.0 & 0.82 & 0.46 & 0.59 \nl
G& 141.4 & 170.5 & 194.9 & 214.8& &38.0 & 53.8 &  56.1 &    - & 0.74 & 0.60 \nl
H&     - & 225.5 & 268.1 & 290.9& &   - & 76.1 &  76.1 &    - &    - & 0.69 \nl
I&     - & 299.1 & 357.3 & 393.3& &   - & 73.6 & 102.5 &    - &    - & 0.77 \nl
J&     - & 375.4 &     - &     -& &   - & 76.3 &     - &    - &    - &    - \nl
\enddata

\tablecomments{
$t_{obs}$=15 $R_b/c$ is representative of the trailing shocks soon after their
production; $t_{obs}$=30 and 60 $R_b/c$ are the times chosen to identify
secondary shocks and emission components. $z$ represents the distance along
the jet axis; $\Delta z$ is the distance from the previous trailing shock; $v$
and $\bar{v}$ are, respectively, the speed and mean speed of the perturbation
in the times and periods considered.\\
$^a$ Observer's time $t_{obs}$ (in $R_b/c$ units) at which the variables are 
measured.}

\end{deluxetable}

\begin{deluxetable}{crrrrrrr}
\tablecolumns{8}
\tablewidth{0pc}
\tablecaption{Observer's frame positions and speeds of the 
              trailing components at selected observer's times\label{tab3}}
\tablehead{
  \multicolumn{1}{c}{ } &
  \multicolumn{4}{c}{$z_{obs} (R_b)$} &
  \multicolumn{1}{c}{$v_{obs} (c)$} &
  \multicolumn{2}{c}{$\bar{v}_{obs} (c)$} \nl \cline{2-5} \cline{7-8} 
  \multicolumn{1}{c}{Comp.} &
  \multicolumn{1}{c}{$15^a$} &
  \multicolumn{1}{c}{$30^a$} &
  \multicolumn{1}{c}{$50^a$} &
  \multicolumn{1}{c}{$60^a$} &
  \multicolumn{1}{c}{$15^a$} &
  \multicolumn{1}{c}{$15-30^a$} &
  \multicolumn{1}{c}{$30-60^a$}}

\startdata

A &   1.0 &   1.5 &   1.9 &   2.2 & 0.16 & 0.03 & 0.05 \nl
B &   4.4 &   5.1 &   4.7 &   4.5 & 0.20 & 0.05 & 0.00 \nl
C &   7.1 &   8.0 &   8.7 &   8.8 & 0.41 & 0.06 & 0.03 \nl
D &   9.8 &  11.4 &  12.8 &  13.3 & 0.53 & 0.11 & 0.08 \nl
E &  12.9 &  16.6 &  18.0 &  18.8 & 0.41 & 0.25 & 0.07 \nl
F &  16.0 &  22.1 &  24.7 &  25.9 & 0.57 & 0.41 & 0.12 \nl
G &  19.6 &  28.0 &  33.7 &  34.7 & 0.41 & 0.56 & 0.22 \nl
H &  22.9 &  36.1 &  44.5 &  50.7 & 0.57 & 0.88 & 0.49 \nl
\enddata

\tablecomments{
$^a$ Observer's time $t_{obs}$ (in $R_b/c$ units) at which the variables are 
measured.}

\end{deluxetable}

\begin{figure*}
\plotone{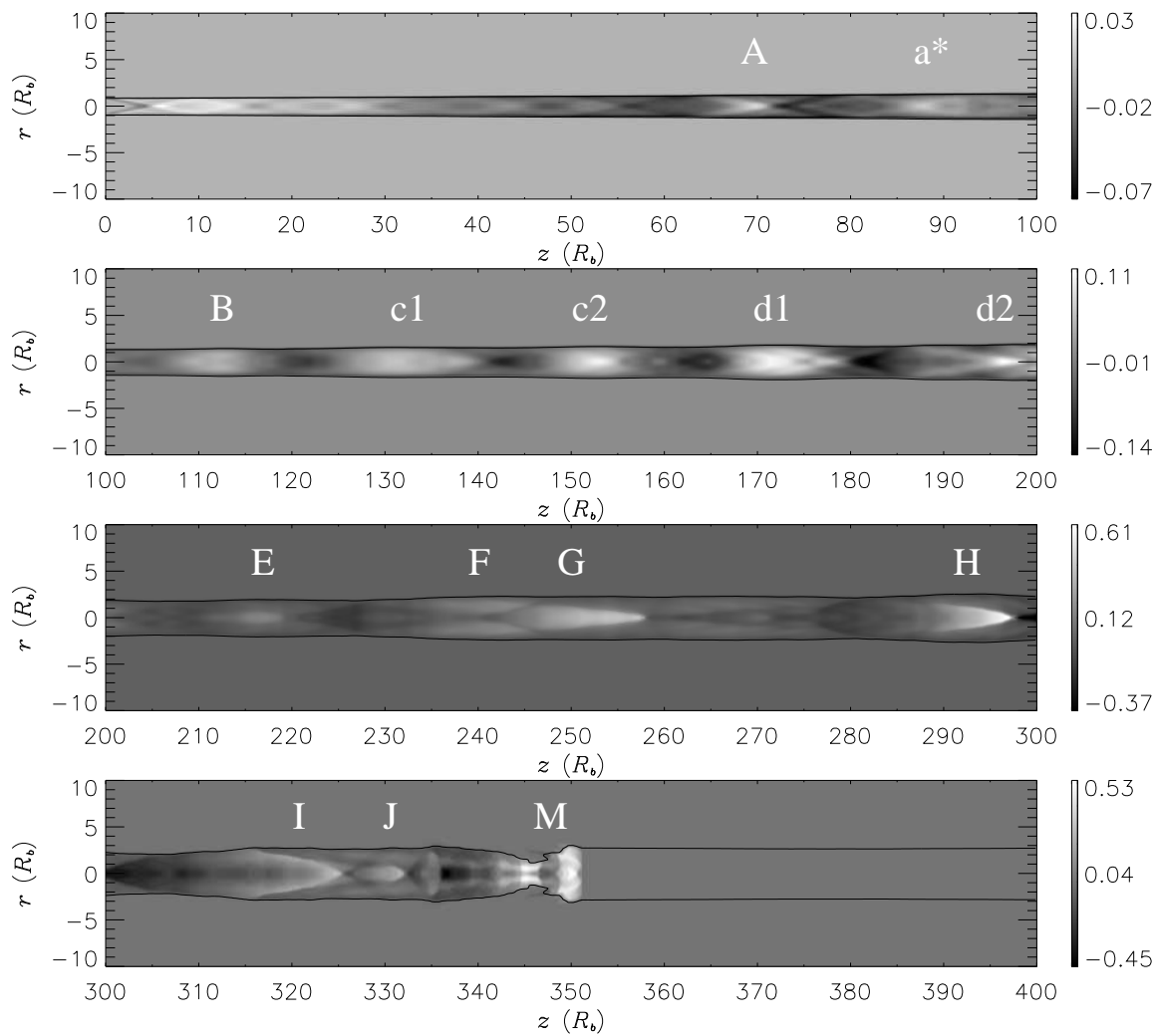}
\caption{Relative variation with respect to the undisturbed steady jet of the
  Lorentz factor (logarithmic scale) at t=350 $R_b/c$. Note the different
  scale ranges in each frame to enhance the representation of the trailing
  conical shocks (labeled A to J) following the main perturbation (M). Typical
  shock angles to the jet axis are $\sim 10-15^\circ$.}
\label{fig1}
\end{figure*}

\begin{figure*}
\plotone{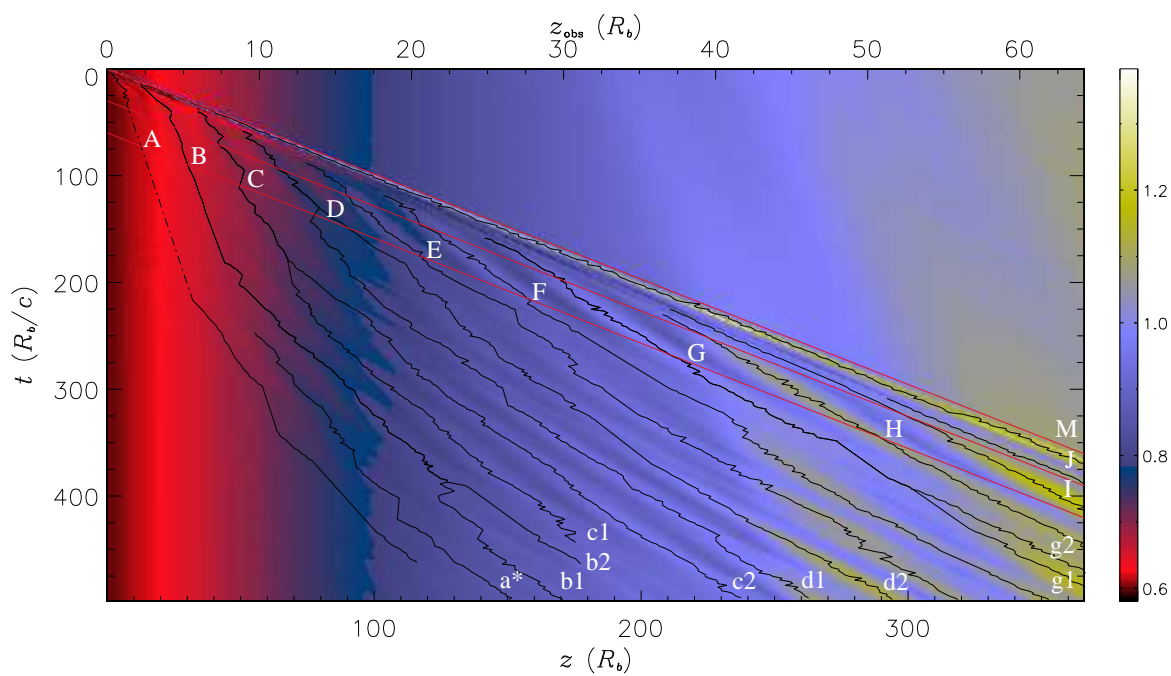}
\caption{Space-time diagram for the logarithm of the jet Lorentz factor
  distribution. The diagonal black line from top left to bottom right marks
  the trajectory of the main perturbation (M), whereas the remaining black
  lines correspond to the tracks of the trailing shocks (identified as local
  maxima at positions where the Lorentz factor is 1\% or more higher than the
  steady jet model). Red lines correspond to constant observer's times
  $t_{obs}$=0, 30, and 60 $R_b/c$ for a viewing angle of $10^{\circ}$.}
\label{fig2}
\end{figure*}

\begin{figure*}
\plotone{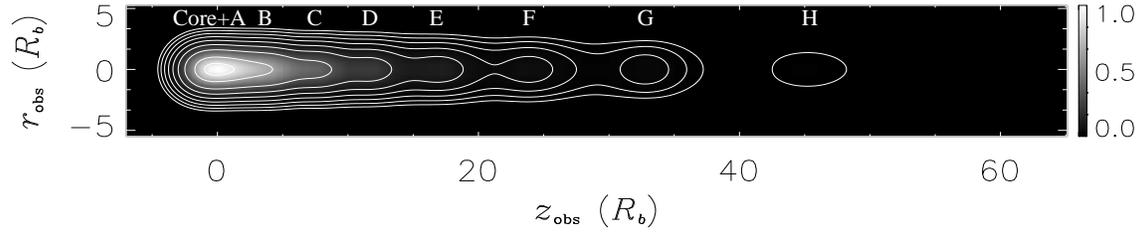}
\caption{Simulated total intensity radio map (convolved with a beam of 2.2
$R_b$ FWHM) for epoch 50 $R_b/c$ in the observer's frame and a viewing angle
of $10^{\circ}$.  Contours are plotted for 0.5, 1, 2, 4, 8, 16, 32, 64, and
90\% of the normalized peak intensity of the core. Components have been
labeled except for I, J and M (associated with the main perturbation) that
have already propagated off the plot to the right.}
\label{fig3}
\end{figure*}

\begin{figure*}
\plotone{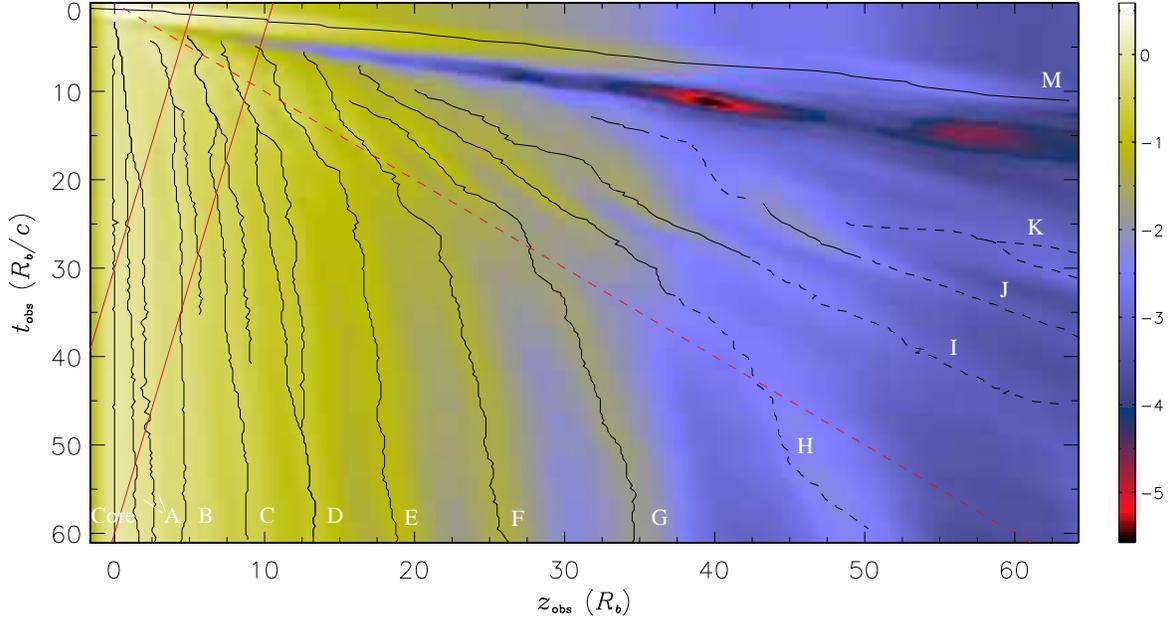}
\caption{Space-time diagram for the logarithm of the mean (unconvolved)
intensity (across slices normal to the jet axis and normalized to the mean
core intensity) in the observer's frame. Black tracks correspond to
trajectories of emission components associated with the main perturbation (M)
and the trailing components. Black dashed lines correspond to components with
intensities below 0.5\% of the normalized value. Red continous lines are
lines of constant time in the source frame (from left to right, $t = 30, 60
R_b/c$). The red dashed line represents a speed equal to $c$.}
\label{fig4}
\end{figure*}

\end{document}